\documentclass[traditabstract]{aa} 
\usepackage{graphicx}
\usepackage{txfonts}
\usepackage{natbib,twoopt}
\usepackage{hyperref} 
\bibpunct{(}{)}{;}{a}{}{,} 
\newcommandtwoopt{\citeads}[3][][]{\href{http://adsabs.harvard.edu/abs/#3}%
{\citealp[#1][#2]{#3}}}
\newcommandtwoopt{\citepads}[3][][]{\href{http://adsabs.harvard.edu/abs/#3}%
{\citep[#1][#2]{#3}}}
\newcommandtwoopt{\citetads}[3][][]{\href{http://adsabs.harvard.edu/abs/#3}%
{\citet[#1][#2]{#3}}} 
\newcommandtwoopt{\citeyearads}[3][][]%
{\href{http://adsabs.harvard.edu/abs/#3}{\citeyear[#1][#2]{#3}}} 

\newcommand{\mic}{$\mu$m}
\def\kms    {\ifmmode{{\rm \ts km\ts s}^{-1}}\else{\ts km\ts s$^{-1}$}\fi}
\def\lsol   {\ifmmode{{\rm L}_{\odot}}\else{L$_{\odot}$}\fi}
\def\msol   {\ifmmode{{\rm M}_{\odot}}\else{M$_{\odot}$}\fi}
\def\hi     {\ifmmode{{\rm H}{\rm \small I}}\else{H\ts {\scriptsize I}}\fi}
\def\hh     {\ifmmode{{\rm H}_2}\else{H$_2$}\fi}

\def\zsol   {\ifmmode{{\rm Z}_{\odot}}\else{Z$_{\odot}$}\fi}
\def\tex {\ifmmode{{T}_{\rm ex}}\else{$T_{\rm ex}$}\fi}
\def\tmb {\ifmmode{{T}_{\rm mb}}\else{$T_{\rm mb}$}\fi}
\begin{document}
\title{A cold gas reservoir to fuel M31 nuclear black hole and stellar
  cluster}


   \author{A.-L. Melchior
          \inst{1,2}
          \and
          F. Combes\inst{1}
          }

 \institute{LERMA,
Observatoire de Paris, LERMA, UMR8112, 61, avenue de l'Observatoire, Paris, F-75
014, France\\
              \email{A.L.Melchior@obspm.fr,Francoise.Combes@obspm.fr}
\and
Universit\'e Pierre et Marie Curie-Paris 6, 4, Place Jussieu,
F-75\,252 Paris Cedex 05, France\\ }

   \date{Received ... ; accepted ... }

 
   \abstract {With IRAM-30m/HERA, we have detected CO(2-1) gas
     complexes within 30\,arcsec ($\sim$ 100\,pc) from the {center
       of M\,31} amounting to a minimum total mass of $4.2\times
     10^4$\,M$_\odot$ (one third of the positions are detected).
     Averaging the whole HERA field, we have shown that there is no
     additional undetected diffuse component.  We show that the above
     gas detection is { associated with} gas lying on the far side
     as no extinction is observed in the optical, but some emission is
     present on infra-red Spitzer maps. The kinematics is complex.
     { (1)} The velocity pattern is mainly redshifted: { the
     dynamical { center} of the gas differs from the black hole
     position and the maximum of optical emission, and only the red-shifted side
is seen in our data.  (2)} Several
     velocity components are detected in some lines of sight. 

     { Our interpretation is supported by the reanalysis of the
       effect of dust on a complete planetary nebula sample. Two dust
       components are detected with respective position angles of
       37\,deg and -66\,deg. This is compatible with the
       superposition of the (PA=37\,deg) disk dominated by the 10-kpc
       ring and the inner 0.7-kpc ring detected in infrared data,
       which position angle (-66\,deg) can be measured for the first
       time. The large scale disk, which dominates the HI data, is
       very inclined (i=77\,deg), warped and superposed on the line of sight
 on the less inclined inner ring. The detected CO emission might come
from both components. 
       }}


   \maketitle
%
 
\section{Introduction}
M31 is usually described as a quiescent galaxy with little star
formation, { at a level of 0.4M$_\odot$yr$^{-1}$}
\citepads[e.g.][]{2006ApJ...650L..45B,2010A&A...517A..77T,2011AJ....142..139A}
and { with ultra-weak nuclear activity} \citepads{2000ApJ...540..741D}.
The presence of a very massive black hole
\citepads{1984ApJ...286...97D} { and the lack of gas within 300\,pc
  \citepads{2006A&A...453..459N,2009ApJ...705.1395C,2009ApJ...695..937B}
  suggest} that the main gas reservoir has been accreted and is
exhausted, { although} some gas is detected within 1\,kpc from the
{ center \citepads{2000MNRAS.312L..29M,2011A&A...536A..52M}}. {
  From optical emission lines} \citetads{1985ApJ...290..136J}
estimated an { ionized} gas mass { on} the order of
1500\,M$_\odot$, which can be accounted for by mass loss from evolving
{ stars.}  \citetads{2012MNRAS.426..892G} rely on Herschel data to
argue that the dust properties are well accounted for by the stellar
heating. { Small amounts of molecular gas have been detected in
  directions} more than 300\,pc from the { center}. { These can
  be associated with} dust features in this area
\citepads{2000MNRAS.312L..29M,2011A&A...536A..52M}.
\begin{table}
\caption{Log of observations}              
\label{table:1}      
\centering                          
\begin{tabular}{c c c c }        
\hline\hline                 
Date & $\langle$ T$_{sys}$ $\rangle$ & t$_{integration}$ & $\#$ pixels (scans) \\    
\hline                        
   8  Nov. 2011 & 320K & 276min   & 36 (144)\\      
   10 Nov. 2011 & 411K & 276min  & 36 (72)\\
   27 Nov. 2011 & 274K & 276min   & 18 (36)\\
   12 Feb. 2012 & 251K & 144min   & 36 (144)\\
   24 Feb. 2012 & 318K & 300min   & 36 (72)\\ 
   11 Mar. 2012 & 351K & 295min   & 36 (72)\\ 
\hline                                   
\end{tabular}
\end{table}
   \begin{figure*}
   \centering
   \includegraphics[angle=-90]{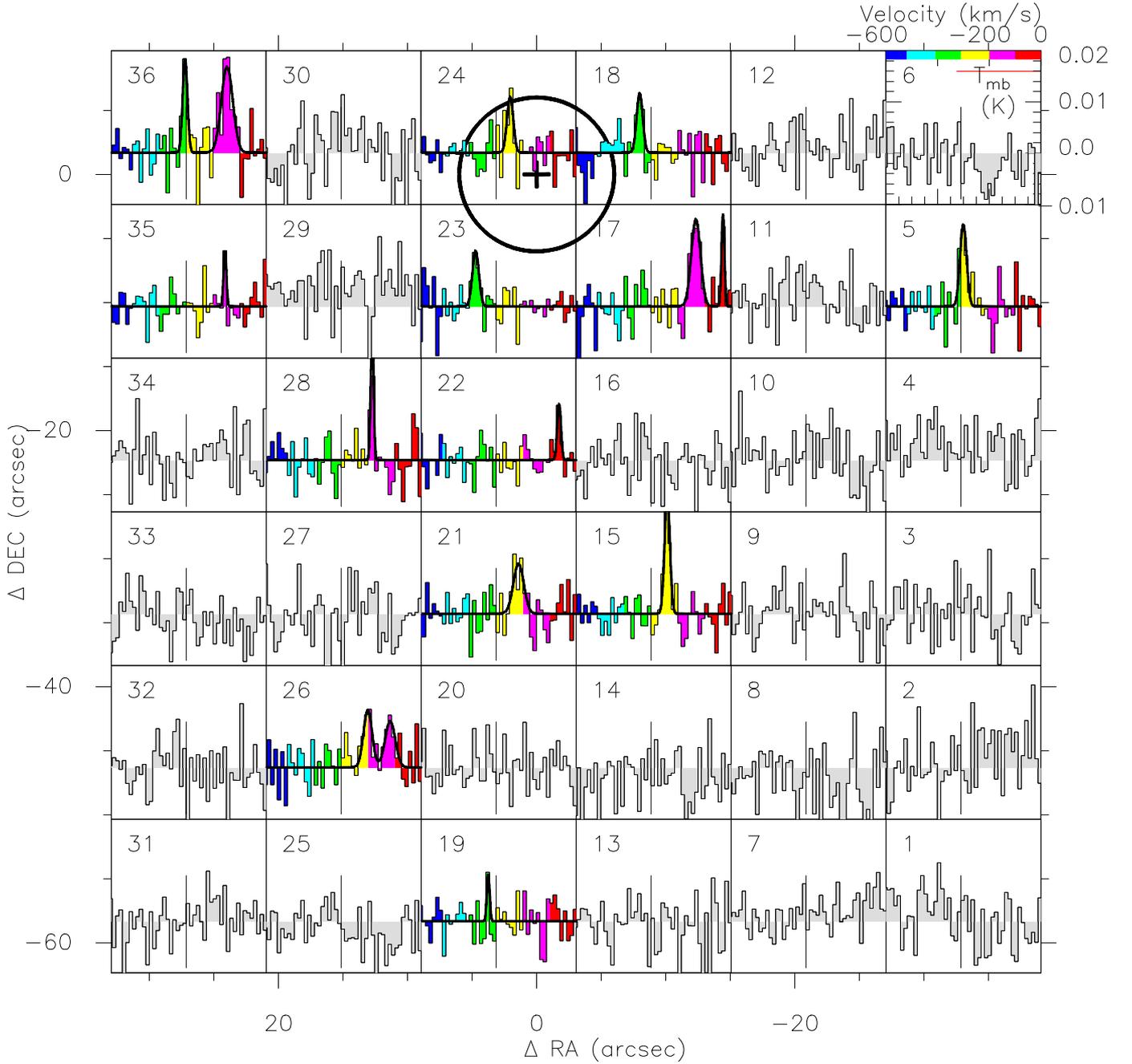}
   \caption{Field M31-1a { centered} on RA: 00:42:44.1 and DEC:
     +41:15:42 (J2000) observed with IRAM-30m/HERA. The { center}
     is indicated with a cross and the { circle} displays the 12''
     beam. The Y scale of each spectrum is in { main}-beam
     temperature (T$_{mb}$ between -0.01 and 0.02\,mK, { as
       indicated on the top right corner}), while the { X-axis
       shows velocity for} the range between -600 and 0\,\kms and
     smoothed to 13\,\kms velocity resolution. A thin line indicates
     the systemic velocity at $-310$\,\kms. { The color coding
       (displayed on the top right corner) of the velocity is used for
       the spectra with $3\sigma$ detections.}}
   \label{FigVibStab}
   \end{figure*}
While the { center} of M31 hosts a supermassive black hole with a
mass of $0.7-1.4 \times 10^8 M_\odot$
\citepads{2001A&A...371..409B,2005ApJ...631..280B}, it is one of the
most silent { ones} \citepads{2010ApJ...710..755G}, although {beginning in 2008} it started to murmur
\citepads{2011ApJ...728L..10L}.  Furthermore, it exhibits many
coherent structures interpreted as tracers of its merging history:
there is a lopsided nuclear disk \citepads{1993AJ....106.1436L} with
two stellar components, P1 and P2 separated by 0.45'' in the {center}. From the kinematics, the black hole is located in between
P1 and P2, but closer to P2.  An A-star cluster \citepads[see
also][]{1999ApJ...522..772K}, detected in a third component (P3) of
M\,31's double nucleus by \citetads{2005ApJ...631..280B}, can be {associated with} a recent star formation episode. { This}
occurred 200\,Myr ago, { involved} a total mass in the range 10$^4$
-- 10$^6 M_\odot$, { and corresponds} to an accretion rate of 10$^{-4}$ --
10$^{-2} M_\odot$\,yr$^{-1}$.  Its presence so close to the black hole
raises a number of issues: how young stars were formed deeply inside
to the tidal field of a supermassive black hole { and} how such stars have
formed while there is no cold gas detected in the { surroundings}
\citepads[e.g.][]{2012ApJ...745..121L,2009MNRAS.397..148L}? In the
Galaxy, SgrA$^*$ has experienced X-ray flares, attributed to the
infall of gas, while a cloud of gas identified by
\citetads{2012Natur.481...51G} is expected to fall onto the black hole
in 2013.  M31$^*$ is experiencing a similar murmur according to
\citetads{2011ApJ...728L..10L} suggesting some gas infall.

Beside the presence of a young star cluster, an ionized gas outflow is
detected in X-rays along the minor axis of the galaxy by
\citetads{2008MNRAS.388...56B}, perpendicular to the main disc.  The
relative intensity of the outflow on both sides is compatible with the
intensity of the observed B extinction: the NW side is more
extinguished than the SE side. As discussed in
\citetads{2011A&A...536A..52M}, the velocity field of the
circumnuclear region (40''$\times$40'' or 150\,pc$\times$ 150\,pc)
measured in optical ionized gas does not exhibit any clear rotation
pattern: beside a spot at the systemic velocity in M31's { center},
the whole area is blueshifted with respect to the systemic velocity.
This coherent flow of ionized gas is decoupled from the stellar
kinematics
\citepads{2005ApJ...631..280B,2000ApJ...540..741D,2010A&A...509A..61S},
and could be connected to the recent star formation activity.  {Please note that we assume throughout the paper a distance to M31 of
  780\,kpc \citepads{2006A&A...459..321V}, i.e.  1\,arcsec$=$3.8\,pc.
  Most up-to-date results, based on cepheids, quote
  752$\pm$27\,kpc \citepads{2012ApJ...745..156R}.  However, for
  coherence with previous works we keep 780\,kpc, which lies within
  $1\sigma$ uncertainties of these new results.}

Some cold gas is expected to feed the black hole, even though { in
  contrast with} the Milky Way, a general lack of HI in the vicinity
of M31's nucleus has been noted for several decades
\citepads{1976MNRAS.176..321E,1982A&AS...49..745B}. In this paper, we
present the first molecular detections within 30\,arcsec from the {center. In Section \ref{sect:obs}, we present the new observations
  performed at IRAM-30m near the center of M31. In Section
  \ref{sect:ana}, we analyze our molecular gas detections and compare
  them with other wavelengths.  In Section \ref{sect:int}, we discuss
  how the interpretation of these data.}.
   \begin{figure*}
   \centering
   \includegraphics[angle=-90,width=6cm]{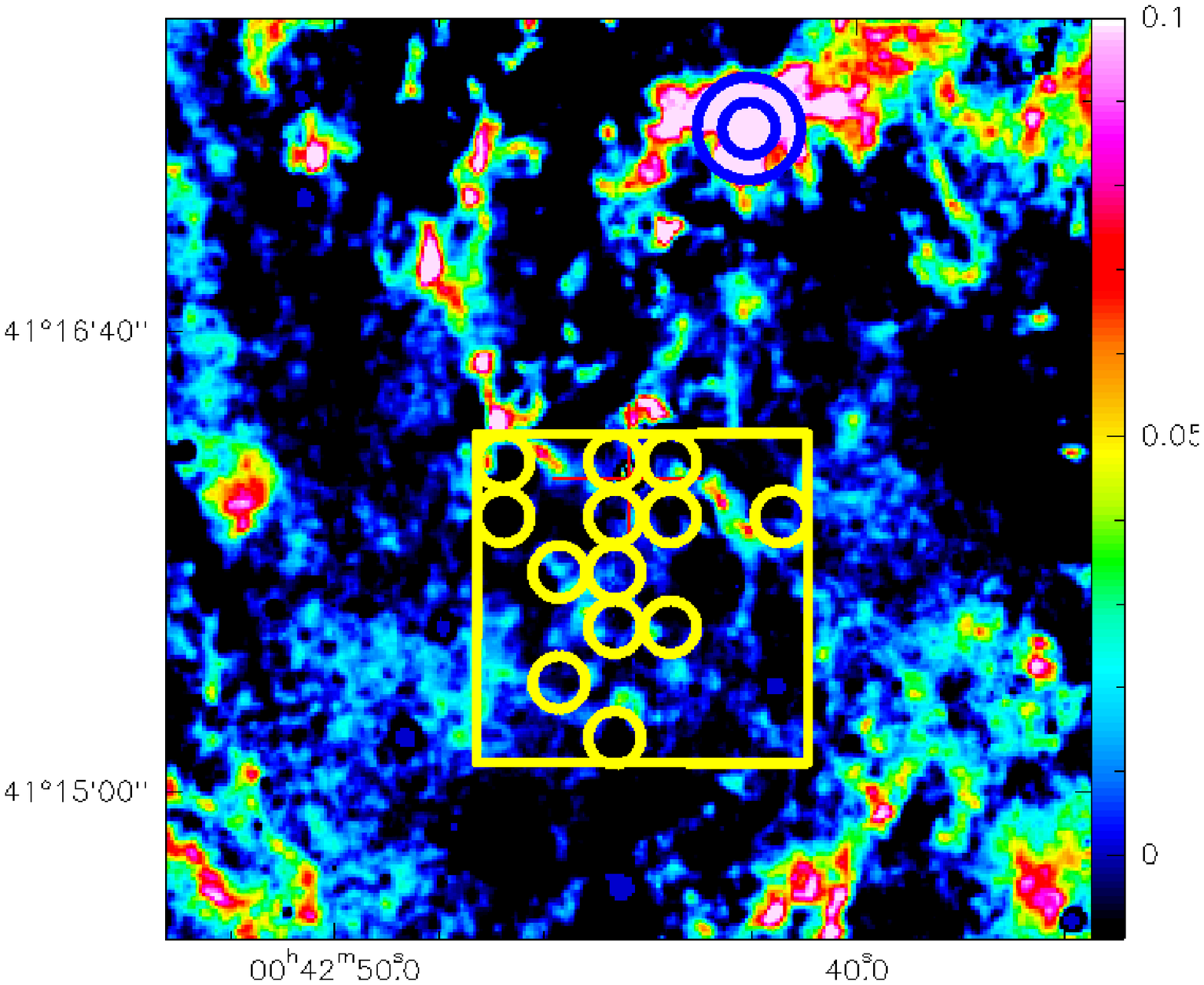}
   \includegraphics[angle=-90,width=6cm]{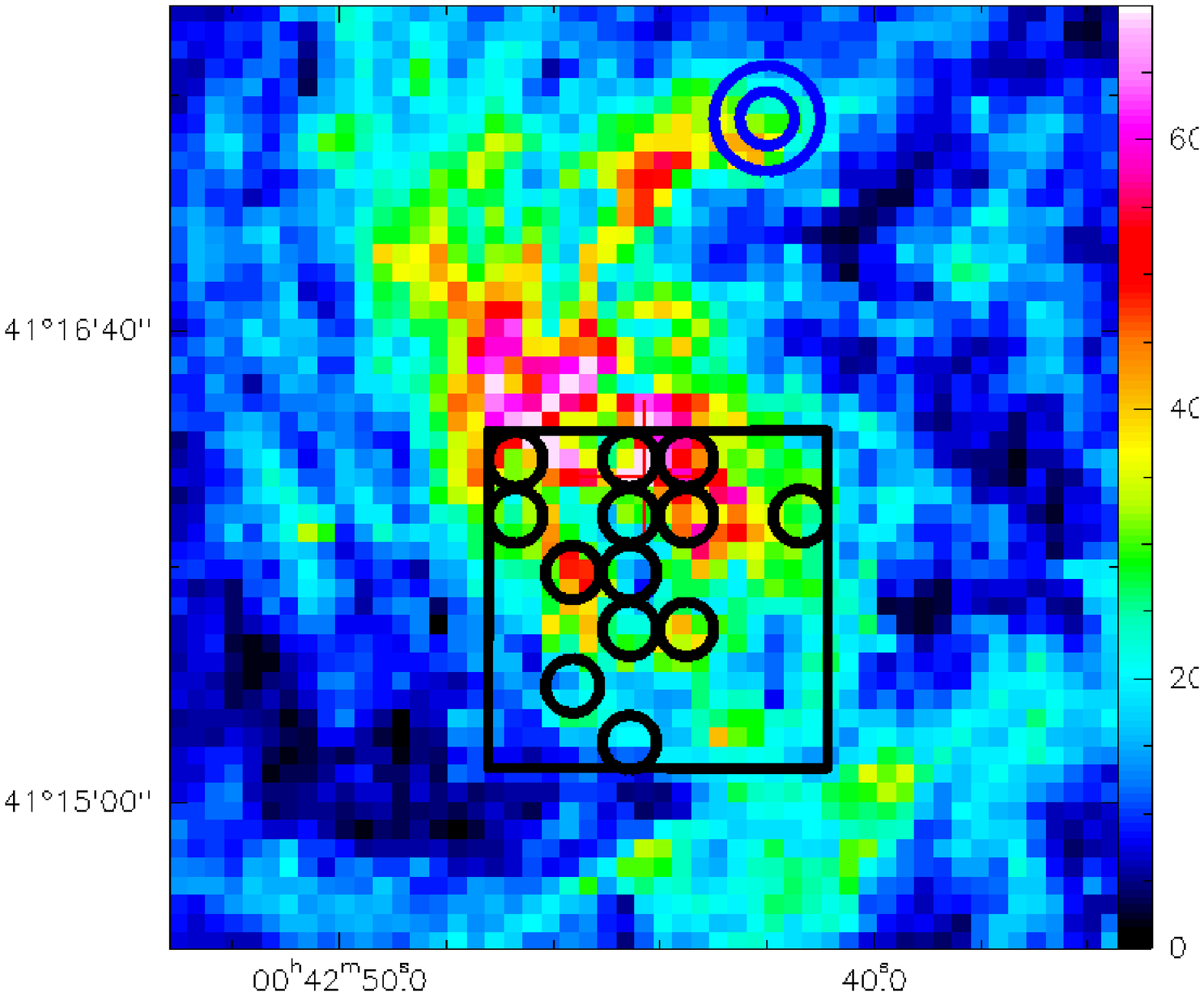}
   \includegraphics[angle=-90,width=6cm]{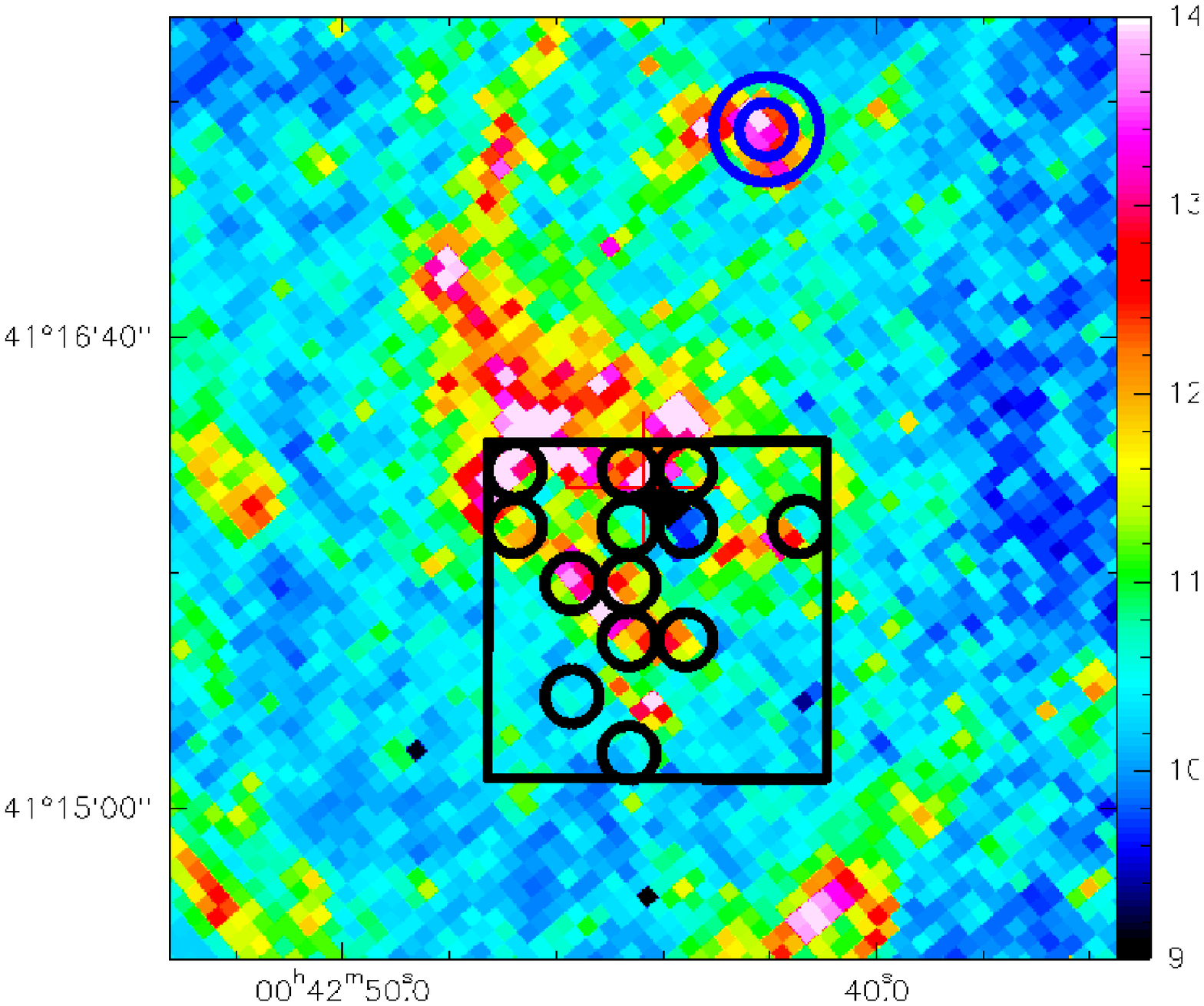}
   \caption{Field M31-1a { centered} on RA: 00:42:44.1 and DEC:
     +41:15:42 (J2000) observed at IRAM-30m HERA, superimposed on the
     A$^B_{observed}$ observed extinction map
     \citepads{2000MNRAS.312L..29M} (left panel), H$\alpha$ and
     $[NII]$ emission map \citepads{1988AJ.....95..438C} (middle
     panel), and 8\mic \,Spitzer emission map
     \citepads{2006Natur.443..832B} (right panel).  The circles
     correspond to the $\ge 3\sigma$ detections discussed in this
     paper.}
   \label{fig:maps}
   \end{figure*}

\section{Observations}
\label{sect:obs}
\begin{table*}[Ht]
  \caption[]{Characteristics of the CO(2-1) lines. Spectra are smoothed to a 13\,\kms resolution, but the fits in italic have been performed on 2.6\,\kms spectra. The offsets refer to the { center} of the M31-1a field.}  
	\label{tab:lines}
        \begin{tabular}{p{0.01\linewidth}p{0.08\linewidth}p{0.04\linewidth}p{0.1\linewidth}p{0.1\linewidth}p{0.1\linewidth}p{0.05\linewidth}p{0.05\linewidth}p{0.08\linewidth}p{0.05\linewidth}p{0.1\linewidth}}
            \hline
            \noalign{\smallskip}
          $\#$     & Offsets & { R ('')}& I$_{CO}$ (K km s$^{-1}$) $=\int T_{mb} dV$& V$_0$ (km s$^{-1}$) & $\sigma$ (km s$^{-1}$) & T$^{\rm peak}_{mb}$ (mK) & baseline rms (mK) & $N_{H_2}$         (cm$^{-2}$) & $\Sigma_{H_2}$ (M$_\odot$\,pc$^{-2}$) & M$_{beam}$ (M$_\odot$)\\
            \noalign{\smallskip}
            \hline
            \noalign{\smallskip}
           5   & -33.0,-8.33 & { 34.0}& 0.51$\pm$0.12 & -301.7$\pm$4.0 & 30.1$\pm$8.6 & 16.0 & 4.2 &$1.17\times 10^{20}$   & 1.99 & $3.25\times 10^{3}$ \\
           15   & -9.0,-32.3 & { 33.5} &0.60$\pm$0.08 & -248.3$\pm$1.7 & 24.4$\pm$3.9 & 23.0 & 3.2 &$1.38\times 10^{20}$   & 2.35 & $3.84\times 10^{3}$ \\
           17   & -9.0,-8.3 & { 12.2}& 0.76$\pm$0.13 & -137.4$\pm$3.9 & 41.9$\pm$7.3 & 17.1 & 4.0 &$1,75\times 10^{20}$   & 2.97  & $4.85\times 10^{3}$ \\
           17   & -9.0,-8.3 & { 12.2} & 0.26$\pm$0.08 & -32.7$\pm$3.9 & 13.3$\pm$29.1 & 18.4 & 4.0 &$0.60\times 10^{20}$   & 1.02  & $1.67\times 10^{3}$ \\
           18   & -9.0,-3.7 & { 9.7} & 0.35$\pm$0.09 & -355.0$\pm$4.1 & 27.8$\pm$7.0 & 11.7 & 3.5 &$0.81\times 10^{20}$   & 1.37  & $2.24\times 10^{3}$ \\
           19   & -3.0,-56.3 & { 56.4} & 0.13$\pm$0.05 & -341.4$\pm$4.0 & 13.1$\pm$3.9 & 9.2 & 3.8 &$0.30\times 10^{20}$   & 0.51  & $0.83\times 10^{3}$ \\
           {\em 19}   & {\em -3.0,-56.3} & {\em  56.4} & {\em
             0.10$\pm$0.03} & {\em -339.9$\pm$4.0} & {\em 7.5$\pm$2.2}
           & {\em 12.9} &{\em  5.1} &{$\mathit  {0.23\times 10^{20}}$}   & {\em  0.39} & {\em  $\mathit  {0.64\times 10^{3}}$} \\
           21   & 3.0,-32.3 & { 32.4 }& 0.50$\pm$0.13 & -224.0$\pm$6.9 & 48.1$\pm$11.1 & 9.8 & 3.9 &$1.15\times 10^{20}$   & 1.96  & $3.20\times 10^{3}$ \\
           22   & 3.0,-20.3 & { 20.5} & 0.20$\pm$0.07 & -67.1$\pm$4.0 & 17.2$\pm$6.2 & 11.0 & 3.0 &$0.46\times 10^{20}$   & 0.78   & $1.27\times 10^{3}$ \\
           23   & 3.0,-8.3 & { 8.8} & 0.34$\pm$0.09 & -390.4$\pm$4.0 & 29.5$\pm$8.1 & 10.9 & 3.2 &$0.78\times 10^{20}$   & 1.33   & $2.17\times 10^{3}$  \\
           24   & 3.0,3.7 & { 4.8} & 0.33$\pm$0.09 & -256.6$\pm$5.0 & 28.6$\pm$8.6 & 11.0 & 3.1 &$0.76\times 10^{20}$   & 1.29    & $2.11\times 10^{3}$ \\
           26   & 15.0,-44.3 & { 46.8} & 0.48$\pm$0.14 & -207.8$\pm$5.7 & 39.8$\pm$14.5 & 11.2 & 4.1 &$1.10\times 10^{20}$   & 1.88& $3.07\times 10^{3}$  \\
           26   & 15.0,-44.3 & { 46.8} & 0.45$\pm$0.16 & -121.1$\pm$7.7 & 47.0$\pm$21.9 & 9.1 & 4.1 &$1.03\times 10^{20}$   & 1.76 & $2.87\times 10^{3}$  \\
           28   & 15.0,-20.3 & { 25.2} & 0.36$\pm$0.08 & -190.9$\pm$1.0 & 13.0$\pm$33.1 & 26.1 & 3.8 &$0.83\times 10^{20}$   & 1.41& $2.30\times 10^{3}$  \\
           {\em 28}   & {\em 15.0,-20.3} & {\em  25.2} & {\em 0.22$\pm$0.05} & {\em -189.7$\pm$1.9} & {\em 14.1$\pm$3.1} & {\em 14.4} &{\em 5.6} & {\em $\mathit  {0.51\times 10^{20}}$}   &{\em 0.86} & {\em $\mathit  {1.40\times 10^{3}}$} \\
           35   & 27.0,-8.3 & { 28.2} & 0.15$\pm$0.05 & -160.9$\pm$3.3 & 13.0$\pm$4.79 & 10.9 & 3.6 &$0.35\times 10^{20}$   & 0.59 &  $0.96\times 10^{3}$ \\
           {\em 35}   & {\em 27.0,-8.3} & {\em  28.2} & {\em 0.064$\pm$0.026} & {\em -155.0$\pm$0.8} & {\em 3.4$\pm$1.8} & {\em 17.8} & {\em 5.4} &{\em $\mathit  {0.15\times 10^{20}}$}  &  {\em 0.25} & {\em $\mathit  {0.41\times 10^{3}}$} \\
           36   & 27.0,3.7 & { 27.3} & 0.94$\pm$0.16 & -152.8$\pm$4.6 & 52.6$\pm$10.6 & 16.8 & 4.1 &$2.16\times 10^{20}$   &3.68   &  $6.01\times 10^{3}$ \\
           36   & 27.0,3.7 & { 27.3} & 0.43$\pm$0.10 & -315.6$\pm$2.8 & 21.9$\pm$5.3 & 18.3 & 4.1 &$0.99\times 10^{20}$   & 1.68   &  $2.74\times 10^{3}$ \\
  \noalign{\smallskip}
            \hline
            \noalign{\smallskip}
\end{tabular}
\label{tab:char}
\end{table*}
\begin{figure}[Ht]
   \centering
   \includegraphics[angle=0,width=0.48\textwidth]{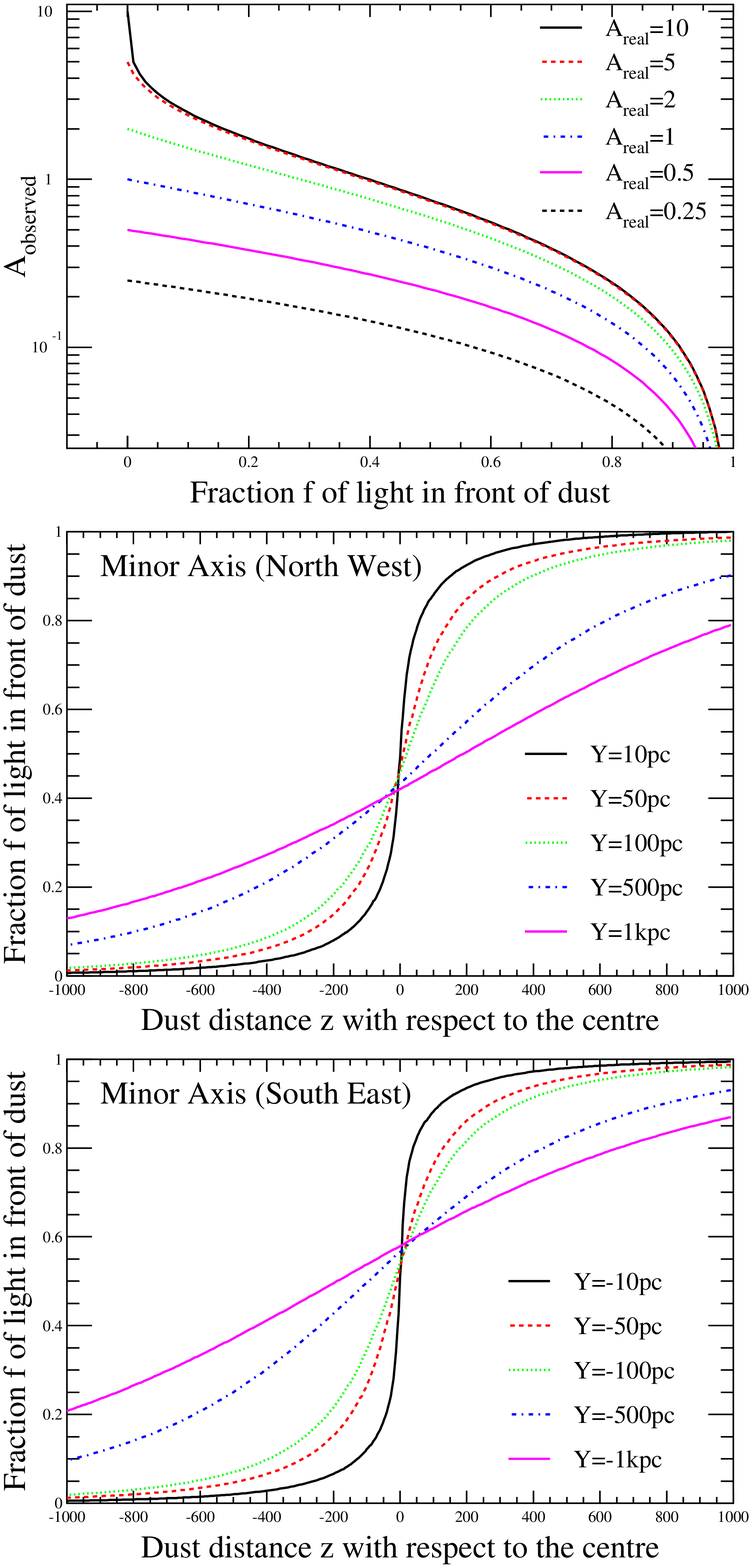}
   \caption{Relationship between extinction and the { position} of the
     dust { clumps} along the line of sight. The { upper} panel
     displays how the observed extinction A$_{observed}$ relates {to} the fraction $f$ of light in front of the dust for
     different values of real extinction A$_{real}$. The { lower}
     panels display how the fraction $f$ relates to the { line of
       sight distance} $z$ with respect to the { center} for
     different { projected distances $Y$ to the { center} along
       the minor axis}. { We detect a near/far side asymmetry,
       which strengthens the effect of absence of observed extinction
       on the far side. We rely} on the { modeling} of
     \citetads{2011A&A...526A.155T}.}
   \label{fig:ext}
 \end{figure} 
   { In the period between November 2011 - March 2012, we used the
     1.3mm multibeam HEterodyne Receiver Array (HERA) at the IRAM-30m
     telescope \citep{2004A&A...423.1171S} to do a CO(2-1) survey of
     M31's 0.7\,kpc inner ring. One of the fields of this survey
     contained the center of M31. We refer to this field as M31-1a and
     show a 60''x60'' map with 12'' angular resolution in Figures
     \ref{FigVibStab} and \ref{fig:maps}.}  We thus performed a 60''
   by 60'' map with 12'' spatial resolution for the CO(2-1) line.
   Data were acquired in wobbler switching mode, using the Wideband
   Line Multiple Autocorrelator (WILMA) facility as backend.  The {wobbler} throw (ranging between 60'' to 210'' in azimuth) has
   been adjusted every { 30\,min, to} avoid extinction areas (and
   possible associated molecular gas emission).  This procedure has
   been followed as best as possible given the constraints of the Pool
   observing periods allocated to this project. Some signal in the OFF
   has been detected in some scans, but as the OFF positions are
   changing with time, { on average it} disappears in the reduction
   process.  Some signal can be underestimated but this should be
   included in the global 20$\%$ calibration uncertainties. A total of
   540 raw spectra were recorded in 21.5\,h of telescope time with a
   spectral resolution of 2.6 \kms. Table \ref{table:1} displays the
   epoch of observations, for each day the average system temperature,
   the total integration time accounting for two independent
   polarization measurements and the number of pixels and scans
   performed. We provide main-beam temperatures throughout this paper
   with beam and forward efficiencies: B$_{eff}$=63 and F$_{eff}$=94.

   Each spectrum was first visually inspected to remove those affected
   by very unstable baselines. { They are then processed in order
     to iteratively correct for the instrumental response: (1) a
     linear baseline is subtracted for each scan, (2) the scans are
     averaged for each position and smoothed. (3)} For the spectra
   where there is a clear detection, a higher order polynomial { is
     subtracted} in order to optimize the baseline subtraction.  The
   typical rms in each final pixel was { at 1$\sigma$ $T_{mb}= 3.9$\,mK}
   in 13 \kms channels. 13 of the 36 positions exhibit a detection
   above 3$\sigma$ (as displayed in Figure \ref{FigVibStab}).

   {\em Upper limit on the continuum.} We have estimated the value of
   the continuum at 1\,mm in the field M31-1a.  4$\%$ of the spectra
     have a mean value outside the range $[$-0.1,0.1$]$\,mK and have
     been removed. The remaining spectra are averaged and smoothed
   over the whole bandwidth (936\,Mhz). We find a continuum level of
   $0.041\pm0.043$\,mK. We thus derive a 3$\sigma$ upper limit of
   0.13\,mK on the continuum emission, corresponding to S$_{\rm
     {continuum}}<0.65$\,mJy.

\section{Analysis}
\subsection{Molecular data}
\label{sect:ana}
Table \ref{tab:lines} summarizes the characteristics of the CO(2-1)
lines detected in this field. A Gaussian function is fitted to each
line to determine its area I$_{CO}$, central velocity $V_0$, width
$\sigma$ and peak temperature T$_{peak}$. The baseline rms is provided
for each line. We assume a { standard} Galactic
X$_{CO}=$N$_{H_2}/$I$_{CO}=$2.3$\times10^{20}$cm$^{-2}$(K\,\kms)$^{-1}$
following \citetads{1988A&A...207....1S}. However, note that different
values have been adopted in the
literature. {\citetads{2011ApJ...737...12L}, relying on
  \citetads{2006A&A...453..459N} CO data with a strong signal
  ($I_{CO}>$1K\,km\,s$^{-1}$), estimate a lower value
  $X_{CO}=9.66\pm1.33\times 10^{-19}$cm$^{-2} (K\,km\,s^{-1})^{-1}$
  for the inner part of M31.}

Relying on the CO(2-1)/CO(1-0) line ratios measured by
\citetads{2011A&A...536A..52M} in this area, we assume a line ratio of
1 and thus adopt the previous X$_{CO}$ ratio for the CO(2-1) line. We
then convert the N$_{H_2}$ column density to an H$_2$ mass
surface density and derive a molecular mass M$_{beam}$ assuming the
gas fills the main-beam. {Lastly,} when all the positions are
averaged a noise level of 0.8\,mK is achieved, but no signal appears.
{This stacking demonstrates that there is no extended emission
  larger than 2.4\,mK.}

One third of the observed positions exhibit a CO(2-1) detection, as
displayed in Figure \ref{FigVibStab}. Beside the spectra 18 and 23
(and the second component of spectra 36), all the detected lines are
redshifted with respect to the systemic velocity.

\subsection{Dust extinction and dust emission}
\label{sect:multi}
{In this area devoid of large amounts of gas
   \citepads[e.g.][]{1996A&A...310...93L}, we have shown in
   \citetads{2011A&A...536A..52M} that in the North-Western part of
   the bulge of M31, CO is detected where extinction is observed,
   while it is not detected in areas where no extinction is measured.
   This supports the dust-gas correlation observed in the Milky Way
   \citepads{1978ApJ...224..132B} and other galaxies
   \citepads[e.g.][]{2012MNRAS.421.2917F}.
   \citetads{2012ApJ...756...40S} is also claiming that the gas in
   M\,31 is well traced by dust at a constant metallicity. The optical
   and near-infrared data displayed in Figure \ref{fig:maps} provide
   complementary information to our CO detection. The left panel
   displays the observed extinction as computed in
   \citetads{2000MNRAS.312L..29M}. The bulge light is mostly dominant
   within $R <~1.2$\,kpc ($\sim 300$\,arcsec) from the center
   \citepads{2011ApJ...739...20C}, where $R$ is the projected distance
   to the center on the sky plane.  We model its photometry with  
   elliptical annuli using the standard surface photometry algorithm
   developed for {\sc IRAF} \citepads{1987MNRAS.226..747J}. This model
   is intended to reproduce the light profile along the bulge of M31
   without extinction. The median intensity over the elliptical
   annulus sectors is used, in order to avoid areas suffering
   extinction. Large scale extinction following the elliptical profile
   could affect this model but it is obviously weak.  The observed
   extinction is defined as $A_\lambda = -2.5 \log_{10}
   (\phi_{obs}/\phi_{model})$, where $\phi_{obs}$ is the observed
   brightness and $\phi_{model}$ the modeled brightness. This provides
   a lower bound of the true internal extinction, with the following
   assumptions: (1) the gas lies in front of the bulge and (2)
   there is only one clump of gas per line of sight. It thus
   eliminates automatically large scale homogeneous extinction like
   the one due to the Milky Way. On the one hand, as the resolution of
   the optical data is close to $ 1$\,arcsec corresponding to 3.8\,pc
   at the distance of M31 (the optical resolution is a factor of 10
   better than the one achieved with CO observations), the second
   assumption is reasonable for most lines of sight given the small
   masses detected.  According to the mass-size relation for molecular
   clouds first observed by \citetads{1987ApJ...319..730S}, we expect
   that the more massive molecular clouds detected here have a size
   smaller than 3\,pc \citep[e.g.][]{2010A&A...519L...7L}. On the
   other hand, it is most probable that all the gas does not lie in
   front of the bulge in the Southern part. The observed extinction
   corresponds to $A_{observed}=-2.5\log_{10}[f+(1-f)\times
   exp({-\tau})]$, where $f$ is the fraction of light in front of the
   dust and $\tau$ the real optical depth at a given wavelength. In
   principle, if several clumps were present along the line of sight,
   we could decompose the extinction $A_{observed}$ into several
   components. In practice, more observational constraints would be
   necessary to perform such a detailed analysis. It is important to
   note that the majority of the dust and associated gas is not
   diluted in the whole bulge and not spread all along the line of
   sight. It is expected to be relaxed and to lie in a disk-like
   structure (or ring). 

   The top panel of Figure \ref{fig:ext} displays how the observed
   extinction relates to the fraction $f$ of 
   light in front of the dust for different values of real internal 
   extinction. For extinction values above 5, it is very difficult to
   disentangle the real value of the extinction from optical
   observations. For a given value of observed extinction, it is,
   however, possible to {put some} constraint on the positions of
   dust clumps.  Relying on the modeling of
   \citetads{2011A&A...526A.155T}, we estimate how the fraction $f$ of
   light in front of the dust relates to the position $z$ of the dust
   clump along the line of sight. The bottom panels of Figure
   \ref{fig:ext} display this relation for different lines of sight
   along the minor axis $Y$ (projected on the sky plane). If the line
   of sight crosses the center (where there is a cusp due to the
   stellar cluster), the fraction $f$ is varying strongly as most of
   the light comes from an area smaller than 1\,arcsec ($\sim
   3.8$\,pc). For lines of sight varying from 10\,pc to 1\,kpc, the
   slope of $f(z)$ is decreasing. There are two main effects to be
   stressed. (1) The observed extinction is very sensitive to the
   location of the dust clump along the line of sight. In the central
   region, dust just behind the mid-plane could easily escape optical
   detection, as the fraction of light in front of the clump can be
   very important. (2) This effect is further strengthened by the
   asymmetry due to the inclination: the far side will have a larger
   fraction of light than the near side.
   
   The real extinction is difficult to  measure from optical data, but
   the observed extinction provides some constraints on the location of 
   the dust along the line of sight as displayed in Figure
   \ref{fig:maps}.  For instance, if $A_{observed}=0.2$, the real
   extinction is larger than 0.25 and the fraction of light $f$ in front
   of the dust between 0.2 and 0.85. Then as a function of the chosen
   line of sight, it is possible to constrain the position of the dust
   clump. For a position at 100\,pc (resp. 10\,pc) from the center, the
   dust clump is expected to lie within 400\,pc (resp. 150\,pc) from the plane
   perpendicular to the line of sight passing through the center. Note
   that these observations have a resolution of 12\,arcsec ($\sim$
   45\,pc), so the gas present very close (in projected distance)
   to the black hole could lie anywhere between 0 and 150\,pc on the
   far side.

   The right panel of Figure \ref{fig:maps} displays the dust emission
   at $8\,\mu$m. It is not affected by extinction, but depends on dust
   grains and their heating. The Southern part exhibits a
   dust-emission intensity much stronger than the observed extinction,
   suggesting that the dust clumps lie on the far side.  }

 \subsection{Characteristics of the dust components} 
\label{sect:pn}
   \begin{figure}[Ht] \centering
     \includegraphics[angle=0,width=0.5\textwidth]{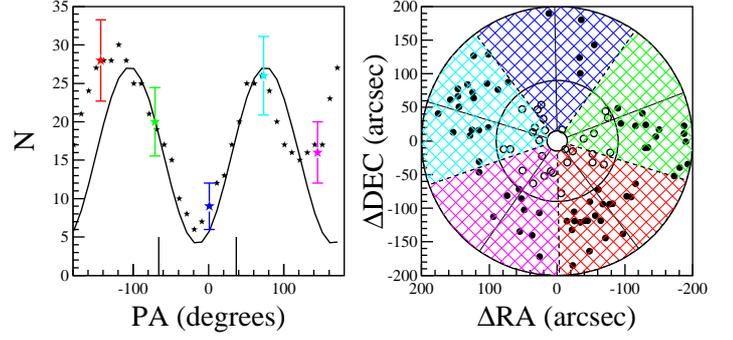}
     \caption{Presence of two dust components affecting the
       \citetads{1989ApJ...339...53C} sample of planetary nebulae. We
       consider a complete sample of 99 planetary nebulae distributed
       into a 15-200\,arcsec annulus. This annulus has been split
       into five equal parts as indicated by the colors. The left
       panel displays the relation between the number $N$ of planetary
       nebulae detected in each position angle. The color star points
       (on the left panel) correspond to the color hatched areas
       displayed on the right panel. The black stars (on the left
       panel) have been obtained with a running sum for different
       position angles (PA).  The error bars indicated only on the
       color points correspond to Poisson statistical noise. The line
       corresponds to the multiplication of two sinusoids of phase
       PA=37\,deg and -66\,deg, as indicated by the ticks at the bottom.
       The right panel displays the spatial distribution of the
       planetary nebula sample in the annuli corresponding to the
       completeness limits (15, 90 and 200\,arcsec).}
     \label{fig:pndust}
   \end{figure} 
   {Following the previous discussion on the expected near/far
     asymmetry expected due to the inclination of M31, one can note
     that \citetads{1989ApJ...339...53C} first detected an asymmetry
     between the near and far sides among the bulge's planetary
     nebulae distribution. A similar geometrical effect is also
     expected in the distribution of microlensing events in M31's
     bulge \citepads[e.g][]{2001MNRAS.323...13K,2006MNRAS.365.1099K}.
     An asymmetry between the near side and the far side, due to
     extinction in the main plane, is observed in the distribution of
     planetary nebulae in the 10\,kpc ring
     \citepads{2006MNRAS.369..120M}.

     We have reinvestigated the catalog of planetary nebulae of
     \citetads{1989ApJ...339...53C}, which samples quite well the bulge area.
     There is a complete sample of 99 planetary nebulae: it is
     spatially complete and \citetads{1989ApJ...339...53C} computed
     the detection efficiencies with respect to the surface magnitude.
     As displayed in the right panel of Figure \ref{fig:pndust}, 29
     are present in a 15-90\,arcsec annulus within a m5007 magnitude
     smaller than 22.1, and 70 in a 90-200\,arcsec annulus with a
     m5007 magnitude smaller than 22.7. This region within 200\,arcsec
     from the center is dominated by the bulge, and the planetary
     nebulae follow the light distribution. In the left panel of
     Figure \ref{fig:pndust}, we have counted the number of planetary
     nebulae in five parts of the 15-200\,arcsec annulus and displayed
     them as a function of the position angle. The five points follow
     a sinusoid. The star points (not independent) have been obtained
     similarly with a running sum for intermediate position angles.

     According to the near/far side asymmetry, one would expect a
     sinusoidal variation of the number of planetary nebulae with a
     2$\pi$ period with respect to the position angle of the main disk
     37\,deg.  Surprisingly, the observed period is $\pi$. The
     overplotted sinusoid is varying as
     $\sin(PA-37)\times\sin(PA+66)$, which is compatible with the
     superposition of two dust components whose main axis have
     respective position angles of 37\,degrees and -66\,degrees. This
     is a new confirmation of the presence of two gas/dust components
     in this region: the main disc (and mainly the 10\,kpc ring) seen
     in projection with a PA of 37\,degree, and the inner ring seen in
     the infra-red (e.g. with Spitzer data) with a position angle of
     -66\,degree. The amplitude of the effect is similar for both components. 
      There is probably an additional perturbation (at the
     limit of detection) close to PA=-180/180\,deg, which might
     correspond to a non-circular structure. More statistics are
     required to be more conclusive.

     Those two dust components expected to be associated with CO
     detections have an orientation compatible with our previous
     discussion: the clumps detected in CO are most probably on the
     far-side. In addition, one could argue that the few points
     detected close to the systemic velocity (and blue shifted) could
     be associated to the main disc (possibly at large scale).}.

 \subsection{Ionized gas} {The middle panel of Figure
   \ref{fig:maps} shows the H$\alpha$ and [NII] emission map
   \citepads{1988AJ.....95..438C}. As discussed by
   \citetads{1971ApJ...170...25R}, it is dominated by [NII] excited by
   shocks. The overall pattern corresponds to the 8\,$\mu$m map, but
   there is not an exact correspondence. (1) These wavelengths are
   affected by extinction. For instance, the position observed in CO
   by \citetads{2000MNRAS.312L..29M} (double circle in Figure
   \ref{fig:maps}) seems affected by extinction.  (2) The kinematics
   of the [NII] line measured by \citetads{1987A&A...178...91B}
   \citepads[see also][]{2011A&A...536A..52M} exhibit a disc in
   rotation and the (40''$\times$40'') circumnuclear region is
   blueshifted with respect to the systemic velocity. In parallel, the
   velocity field measured in CO is redshifted. It is thus probable
   that both components are decoupled. 

   This can be related to the compilation of all gas velocities in the
   inner 10\,arcmin of M31 by \citet{1994ApJ...426L..31S}. The
   isovelocity curves are very irregular and chaotic, even involving
   much larger scales than here.}

\subsection{Bulge light emission}
\begin{figure}[Ht]
   \centering
   \includegraphics[angle=0,width=0.5\textwidth]{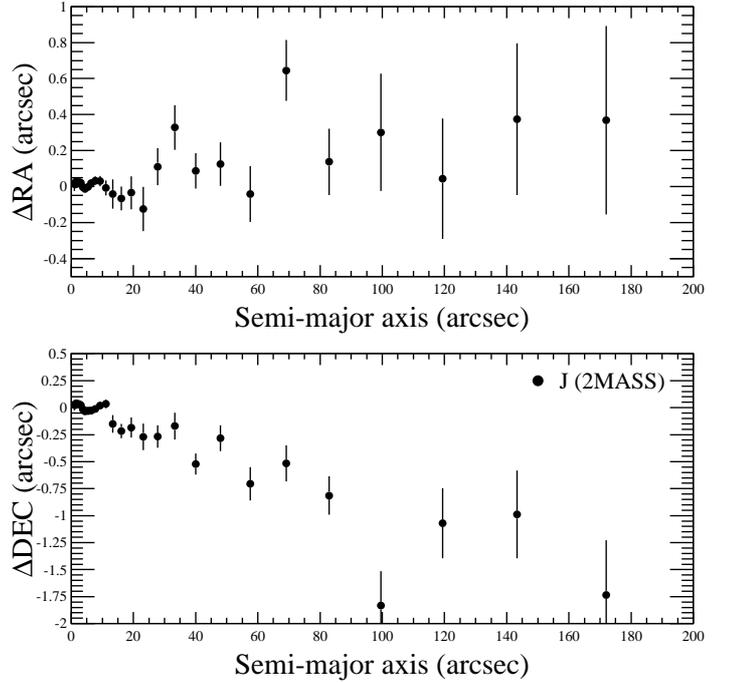}
   \caption{Variations of the position of the centers of the
     elliptical annuli computed on the 2MASS J image as a function of
     the semi-major axis.}
   \label{fig:center}
 \end{figure} {The bulge model computed on photometric images as
   described in Section \ref{sect:multi} provides the ellipse geometry
   parameters \citepads[see e.g.][]{1989AJ.....97.1614K}. We performed
   this modeling on 2MASS J data\footnote{Note that the J image has
     been chosen as it offers a better signal-to-noise ratio than the
     K image.}  \citepads{2006AJ....131.1163S} and Figure
   \ref{fig:center} displays the position of the centers of each
   annulus computed. In contrast with B and H$\alpha$/[NII] data, we
   do not expect any bias due to dust obscuration.  Interestingly, the
   center of the annuli is systematically shifting towards the South
   by about 7\,pc within 0.78\,kpc. It is tempting to compare this
   off-centering of the bulge with the off-centering ($\sim\,350$\,pc)
   of the inner dust ring detected by \citetads{2006Natur.443..832B}.
   Under the hypothesis of a coupled $m=1$ motion between the inner
   bulge and the disk, and given the \citetads{2012A&A...546A...4T}
   estimation the mass of M31's bulge of the order of $4\times 10^{10}
   M_\odot$, we can expect a mass of gas in the inner ring of $9\times
   10^8 M_\odot$, if the maximum of light corresponds to the
   barycenter.

   As discussed in the Appendix \ref{sect:triax} and shown in Figure
   \ref{fig:pa}, the bulge is triaxial but we do not expect our
   results to be affected as the amplitude of the twist is about
   10\,deg.}
 
\section{Interpretation}
\label{sect:int}
As displayed in Figure \ref{fig:maps}, some detections (7/13)
correlate well with the Spitzer dust emission and several
positions\footnote{The Spitzer map seems to exhibit a defect close to
  the {center} at the position of spectrum 17.} (5, 17, 18, 19, 23
and 26) do not. {None of the positions detected in CO correspond
  to any} observed A$^B_{observed}$ extinction. {According to the
  available kinematics, it is most probable that the ionized component
  is decoupled from the molecular gas. The} observed extinction is
smaller than 0.025 in B (see left panel of {Figure \ref{fig:maps}).
Relying} on Figure \ref{fig:ext}, we can derive that the {typical}
fraction $f$ of light in front of the average clump is larger than
{90$\%$}.  Following the {modeling} of
\citetads{2011A&A...526A.155T}, the average clump lies at a depth
between {20 and 200\,pc} from the {center} on the far side,
depending on its {projected distance}. It could be further if the
real extinction is significantly larger than 0.25. {Accordingly,}
the positions with no Spitzer infrared emission could be much further
where the radiation field is too weak to heat the dust, except within
{4\,pc} of the black hole, where the light of the nuclear star cluster
{prevents} the detection of any extinction.

The kinematics is complex and do not exhibit a clear pattern. Most of
the lines are redshifted with respect to the systemic velocity while 4
are blueshifted. This is surprising as the opposite trend was observed
in this field by \citetads{2011A&A...536A..52M} with the optical
ionized gas.  The {gas detected} in the North-East side
\citepads{2000MNRAS.312L..29M} was also redshifted, so it is not
simply a counter-rotation. In addition, the velocity range is spread
between -33 and -390\,\kms.

These results are compatible with the 0.7\,kpc inner ring scenario
discussed in \citetads{2011A&A...536A..52M} and initially proposed by
\citetads{2006Natur.443..832B}: the ring is tilted, which could
explain why the gas is here on the far side and the velocities
redshifted. It {lies off-center}, which could explain why we do
not see a regular rotation pattern, {since we are far from the
  kinematical center}. {As supported by the dust components
  detected with the planetary nebulae in Sect. \ref{sect:pn}, the 0.7-kpc
  inner ring is most probably superimposed on the 10-kpc ring in the
  main disc: this could account for the multiple velocity components
  together with the clumpiness.}

\vspace{0.3cm}

In summary, we have shown the presence of molecular gas close to the
black hole.  There is no extended diffuse molecular emission but we
have detected small dense clumps located on the far side of the bulge.
The detected clumps are located {between 20\,pc and 215\,pc} in
projected distance from the {center} {(and observed with a
  resolution of 45\,pc).} Assuming a single dust/gas clump per line of
sight {and some modeling assumptions}, we show that {clumps
  corresponding to these lines of sight lie} on the far side, at least
at {150\,pc} in depth from the {center} and {most probably
  closer than 600\,pc}. { If some} gas { is present} next to the
{ center, it is also most probably} on the far side but could be
very close to the black hole.  The kinematics exhibit an unexpected
trend: most of the lines are redshifted with respect to the systemic
velocity, which could be due to the off-{ centering}.  Along
several lines of sight, there are multiple CO components. { Our
reanalysis of the \citetads{1989ApJ...339...53C} catalog of planetary
nebulae reveals the presence of two components of position angles
37\,deg and -66\,deg. In the light of our molecular detections
discussed here, we claim that dust and gas from the outer 10-kpc ring
superimpose on the inner 0.7-kpc ring in this very central region. The
inclination of both components is such that the gas and dust in the
Southern area lie mainly on the far side.}

\begin{acknowledgements}
  This article is based on observations (067-11, 221-11) performed in
  the IRAM-30m Pool session at Pico Veleta (Spain). We are most
  grateful to Manuel Gonzalez who monitored this program through the
  IRAM-30m Pool observing sessions. We also thank the IRAM staff at
  Pico Veleta for excellent support at the telescope. IRAM is
  supported by INSU/CNRS (France), MPG (Germany) and IGN (Spain). We
  thank the anonymous referee for his constructive comments. David
  Valls-Gabaud is warmly acknowledged for his helpful comments. This
  publication makes use of data products from the Two Micron All Sky
  Survey, which is a joint project of the University of Massachusetts
  and the Infrared Processing and Analysis Center/California Institute
  of Technology, funded by the National Aeronautics and Space
  Administration and the National Science Foundation.
\end{acknowledgements}

\bibliographystyle{aa}    


\appendix

\section{Triaxality of the bulge}
\label{sect:triax} 
 { }
   \begin{figure}[Ht]
   \centering
   \includegraphics[angle=0,width=0.5\textwidth]{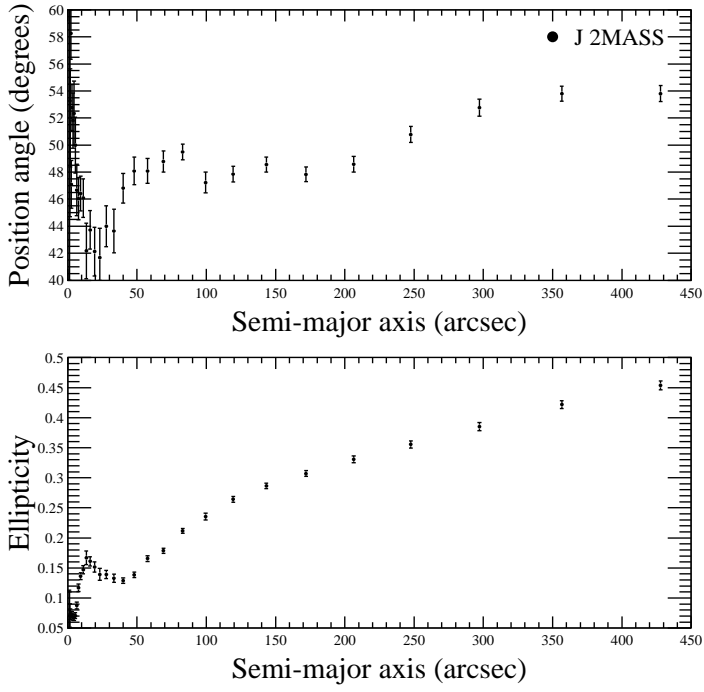}
   \caption{ Position angles and ellipticities of the centers of
     the elliptical annuli computed on the 2MASS J image as a function
     of the semi-major axis. These points correspond to the
     centers of the annuli presented in Figure \ref{fig:center}.}
   \label{fig:pa}
 \end{figure} { Figure \ref{fig:pa} displays the variation of the
   position angles and ellipticities computed in the modeling
   described in Sect. \ref{sect:multi}. There is a clear isophot twist
   which is not due to extinction. The most plausible explanation is
   the triaxiality of the bulge as discussed by
   \citetads[e.g.][]{1982modg.proc..113K}.  Our values are similar to
   those presented by \citetads{2007AAS...21110417B}, but they are
   presented here in linear scale in accordance with Figure
   \ref{fig:center}.

   The isophot twist is significant but its amplitude does not exceed
   10\,deg and should not affect significantly our results. (1) The
   bins used to compute the position angle of the dust component from
   the planetary nebulae distribution are 72\,deg. (2) This
   triaxiality could affect the near/far side effect, but it should be
   a second order effect. }

\end{document}